\documentclass[aps,prl,twocolumn,showpacs,superscriptaddress,preprintnumbers,amsmath,amssymb]{revtex4}
\usepackage{amsmath}
\usepackage{graphicx}
\usepackage{dcolumn}
\usepackage{bm}
\usepackage{float}
\usepackage{xcolor}
\begin{document}
\preprint{\textit{Preprint}}
\title{Anomalous Effects in Single-slit Diffraction of Light at Relativistic Intensities}
\author{Longqing Yi}
\thanks{lqyi@sjtu.edu.cn}
\affiliation{Tsung-Dao Lee Institute, Shanghai Jiao Tong University, Shanghai 201210, China}
\affiliation{Key Laboratory for Laser Plasmas (Ministry of Education), School of Physics and Astronomy, Shanghai Jiao Tong University, Shanghai 200240, China}
\affiliation{Collaborative Innovation Center of IFSA (CICIFSA), Shanghai Jiao Tong University, Shanghai 200240, China}

\date{\today}


\begin{abstract}
High-order harmonic generation via single-slit diffraction of relativistic laser pulses is investigated. 
Using fully kinetic 2D and 3D particle-in-cell simulations, we show that interesting optical phenomena emerge, including the generation of harmonic beams that are anomalously polarized orthongal to the driver, harmonic spectrum variation depending on the incident laser polarization, and a deflection of transmitted lights that leads to a tilted harmonic intensity pattern.
It is shown these anomalous effects are associated with complex peripheral electron dynamics on the plasma vacuum interface. To account for these effects, a new theoretical model is developed to calculate harmonic fields from arbitrary 2D electron motion within the diffraction plane, the results agree well with the simulations.
Our model indicates that the optical properties of the harmonic beams are determined by the 2D in-plane electron motion, which can be leveraged to provide immense opportunities for manipulating light-matter interaction at relativistic intensities.

\end{abstract}

\pacs{}
\maketitle

With the invention of the chirped pulse amplification \cite{CPA}, the laser-matter interaction enters the relativistic regime, 
where the electron quiver motion in the laser fields approaches the speed of the light, giving rise to abundant relativistic nonlinear effects \cite{Mourou2006}. 
For example, the reflection of a relativistically-strong laser can lead to high-harmonic generation (HHG) by the ``relativistic oscillating mirror (ROM)" mechanism \cite{Bulanov1994,Lichters1996,Baeva2006}. 
When the laser beam carries orbital angular momentum (OAM) \cite{Allen1992,Shi2014,Vieira2016}, the reflection converts the OAM from the driver to the harmonics \cite{Zhang2015,Denoeud2017}, and a deflection effect is found at oblique incidence due to the asymmetric radiation pressure of an intense vortex beam \cite{Zhang2016}. 
These behaviors of light not only provide fundamental insight into relativistic optics, but also promise new advances in a variety of applications, such as attosecond sciences \cite{Krausz2009,Heissler2012,Pirozhkov2006,Ma2015}, optical communications \cite{Wang2012,Gibson2004}, and biophotonics \cite{Willig2006}.


On the other hand, another fundamental optical process, the diffraction of light, remains largely unexplored at relativistic intensities. This is probably due to the prepulses tend to destroy small diffraction structures.
Thanks to the progress of laser cleaning technology \cite{Thaury2007,Jullien2006}, the laser temporal contrast has been steadily improving. 
To date, multiple diffraction effects are reported experimentally using self-induced plasma apertures\cite{Gonzalez2016NP,Gonzalez2016NC,Duff2020}. Thus, developing a predictive model for the relativistic light diffraction is 
critical for interpreting future experiments. 
Recently, it is found the diffraction of relativistic laser beams also produce high-order harmonics\cite{Yi2021,Trines2024,Jirka2021,Bacon2022}. Analogous to the ROM mechanism, a ``relativistic oscillating window (ROW)" model has been proposed \cite{Yi2021}, which attributes the HHG to the relativistic Doppler effect due to laser-driven peripheral electron oscillations on the rim of a diffraction window.
 
Despite the similarity, a unique feature of the ROW is that the electron motion that leads to HHG is intrinsically two-dimensional (2D), which 
potentially enables vast opportunities for controlling laser-matter interaction. 
For instance, a circularly polarized laser drives chiral electron oscillation, and this chirality can be imprinted to the harmonics, facilitating spin-orbital interaction of light \cite{Yi2021}. By tailoring the shape of the aperture, this scheme enables independent control over the spin and orbital momentum of harmonic beams \cite{Trines2024}. In addition, the electron acceleration associated with the relativistic diffraction can be harnessed to produce isolated sub-femtosecond electron bunches \cite{Hu2024}.
However, to the best of our knowledge, previous works mainly focus on diffraction apertures with rotational symmetry \cite{Yi2021,Trines2024}, because of the potential to generate harmonic vortices. 
And the ROW model assumes the aperture does not change shape (``rigid window"), 
making it impossible to capture the full complexity of the 2D dynamics of diffraction window.

In this letter, 
we show that, without losing the fidelity of HHG, the restriction on ``rigid window" can be lifted by taking into account only the contribution within a narrow band at the edge of the diffraction window.
Such a quasi-1D structure can be twisted to account for arbitrary 2D electron dynamics.
Using the approach, single-slit diffraction of a relativistic laser pulse is studied and several anomalous effects are identified for the first time, including the generation of harmonic beams polarized orthogonal to the driver, HHG spectrum variation determined by the incident pulse polarization, and a deflection of harmonics due to total angular conservation of light.\\

\begin{figure}[!t]
\centering
\includegraphics[width=8.5cm]{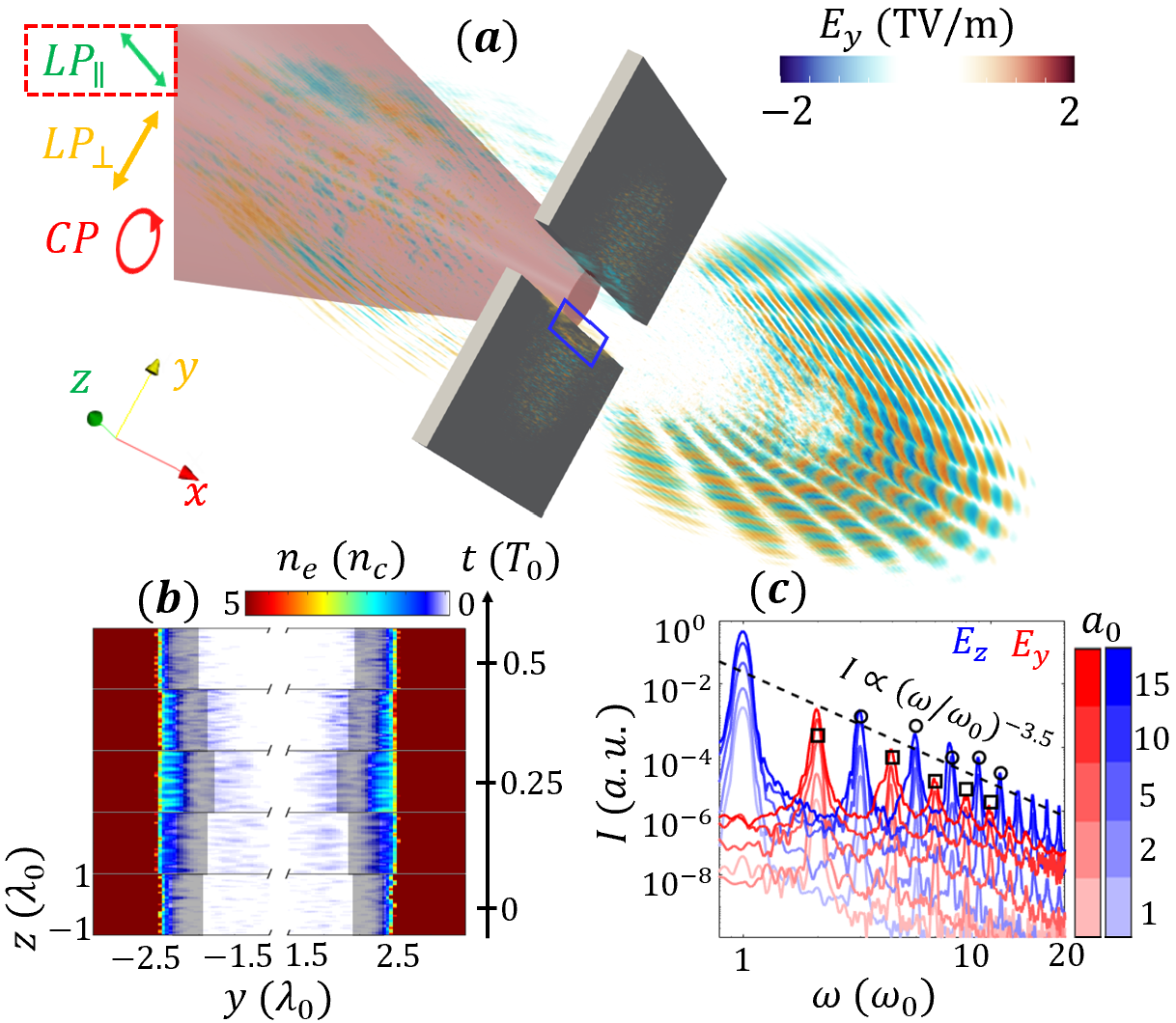}
\caption{(a).Sketch of single-slit diffraction of relativistically strong laser pulses, with three polarization state under consideration (LP$_\parallel$, LP$_\perp$, and CP). 
The color-coded diffracted field in (a) shows the anomalously polarized harmonic fields $E_y$ driven by a LP$_\parallel$ driver, where
(b) shows the electron density fluctuation over a full oscillation cycle ($\sim 0.5T_0$) for this case. The dark banded region represents the area over which the Kirchhoff integral is calculated to yield the harmonic fields.
The HHG spectra are presented in (c), where the blue and red colors show the odd- and even-order harmonics that polarized in $z$ and $y$ directions, respectively. The strength of the colors represents the LP$_\parallel$ driver intensity. The black circular and square markers in (c) show the saturated spectrum calculated from Eq.~(1).}
\label{fig:1}
\end{figure}

The relativistic single-slit diffraction simulation setup is sketched in Fig.~1(a): a high-power laser beam propagating along $x$-axis, irradiating normally on a thin foil with a single slit along the $z$-axis. 
The simulations are performed with particle-in-cell (PIC) code {\sc epoch} \cite{Arber2015}. The laser profile used in the simulation is $\mathbf{E}_l = (\sigma_y\mathbf{e}_y + i\sigma_z\mathbf{e}_z)/\sqrt{\sigma_y^2+\sigma_z^2}E_0\sin^2(\pi t/\tau_0)\exp[-(y^2+z^2)/w_0^2]\exp(ik_0x-i\omega_0 t)$, for $0<t<\tau_0 = 54$ fs, where $\mathbf{e}_y$ ($\mathbf{e}_z$) is the unit vector in $y$ ($z$) direction, $E_0$ is the laser amplitude, $w_0 = 5  {\rm \mu m}$ the focal spot size, $k_0 = 2\pi/\lambda_0$ the wavenumber, and $\lambda_0 = 1 {\rm \mu m}$ the wavelength.
The polarization of drive laser pulse is predominant in determining peripheral electron dynamics, which is controlled by $\sigma_y$ and $\sigma_z$ in the simulation. Three polarization are considered in this work: linear polarization parallel (LP$_\parallel$, with $\sigma_y = 0,~\sigma_z=1$) and perpendicular (LP$_\perp$, with $\sigma_y = 1,~\sigma_z=0$) to the slit, as well as circular polarization (CP, with $\sigma_y = 1,~\sigma_z=\pm1$), as illustrated in Fig.~1(a). The default normalized laser amplitude is $a_0 = eE_0/m_{\rm e}\omega_0c = 5$ for LP$_\perp$ and CP casae, and $10$ for the LP$_\parallel$ case, unless otherwise specified.
The thin foil target [assumed aluminum (Al)] is modeled by a pre-ionized plasma with a thickness of $0.2~{\rm \mu m}$, and electron density $n_0 = 30n_{\rm{c}}$, where $n_{\rm{c}} = m_{\rm{e}}\omega_0^2/4\pi e^2 \approx 1.1\times 10^{21}$ cm$^{-3}$ is the critical density, where $e$, $m_{\rm{e}}$, and $\omega_0$ denote the elementary charge, electron mass, and the laser frequency, respectively
The half-width of the slit is $l_s = 2.5 {\rm \mu m}$, placed at $x_0 = 2\mu m$, with a density gradient at the boundary $n_e(y) = n_0\exp[(|y|-l_s)/h]$ for $|y|<l_s$, where $h = 0.2 {\rm \mu m}$ is the scale length.
Throughout this study, we use 2D simulations to study the HHG spectra, which is performed with a high resolution $dx\times dy = 10 {\rm nm}\times 20 {\rm nm}$ with 50 macro particles for electrons and 10 for Al$^{3+}$ ions per cell; 
and 3D simulations are relied on to capture the full electron dynamics on the diffraction screen, a reduced resolution $dx\times dy \times dz = 40 {\rm  nm}\times 50 {\rm  nm}\times 50 {\rm  nm}$ is used for computational efficiency, with 5 macro particles for electrons and 3 for Al$^{3+}$. In both cases, a high-order particle shape function is applied to suppress numerical self-heating instabilities \cite{Arber2015}.


We first discuss the anomalous harmonic fields generated by a LP$_\parallel$ driver, as represented by the color-coded 3D electric field in Fig.~1(a), with an amplitude reaches 2~TV/m, approaching the relativistic threshold. 
Interestingly, this anomalous diffracting field is orthogonally polarized to the incident laser pulse, and contains only even-order harmonics (Fig.~1(c)).

To understand this effect, we examine the peripheral electrons motion perpendicular to the boundary.
For a LP$_\parallel$ driver, the laser field perpendicular to the slit vanishes, the electron motion is dominated by the ponderomotive force. 
Consequently, the peripheral electron density oscillate at $2\omega_0$ as shown in Fig.~1(b).
Since the ponderomotive force have opposite signs at the $\pm y$ side, the width of the slit varies significantly. Therefore the ``rigid window" assumption in the ROW model is no longer valid in this case.
 
Nevertheless, because the harmonic beams are generated by the nonlinear interaction between the drive laser and the oscillating peripheral electrons, it only occurs within a narrow area near the edge of the diffraction window.
The central region, where the drive pulse simply transmitted though, only affects the fundamental harmonic, which can be safely neglected. The HHG fields can thus be obtained from the Kirchhoff integral \cite{Yi2021}
\begin{equation} 
\begin{split}
\mathbf{E}&_{\rm hhg}(x,y,z,t) = \mathbf{E}(x,y,z)\exp(-i\omega_0t) \\
&= \frac{1}{2\pi}\nabla\times\int_{\rm B}(\mathbf{e_n}\times \mathbf{E_0})\frac{\exp{[ik_0R'-i\omega_0t]}}{R'}ds'.
\end{split}
\label{eq1}
\end{equation}
where the integration is over a narrow banded area (B) near the boundary of the slit. 
The unit vector $\mathbf{e_n}$ is normal to the screen, and $R' = |\mathbf{R}-d\mathbf{R'}|$ is the distance between an observer at $(x,y,z)$ and elementary source [$ds'(x_0, y',z')$], measured at retarded time $t' = t - R'/c$. Here $\mathbf{R}$ is the initial distance, and $d\mathbf{R'}$ denotes the shift of $ds'$ in the diffraction screen.\\

In the following, we apply this method to compute the harmonic spectra produced by a LP$_\parallel$ drive pulse. The width of the integration band is chosen to be $\lambda_0/4$, corresponding to the maximum distance the electrons can travel in half of its oscillating period. 
The shift of elementary sources in the integral band is determined by the peripheral electron dynamics, which is very complicated due to collective effects in plasma. Fortunately, the harmonic beams are mostly generated in the ``surface wake breaking (SWB) phase" (discussed below), where the oscillating surface electrons gain enough energy from the laser that they escape from the plasma \cite{Tian2012,Thevenet2015,Yi2019}. During this period, the electron dynamics is relatively simple, resembling single electron motion in relativistic laser field, the plasma collective effects are not important.
In the SWB phase, the temporal shift of $ds'$ can be approximated by $d\mathbf{R'}(t') = \mp \delta_2\sin(2\omega_0 t')\mathbf{e}_y$, where $\delta_2$ is the amplitude, and the $\mp$ sign corresponding to the opposite phase at the $\pm y$ side of the slit, respectively. 
One can then solve for the retarded time and distance numerically to yield the harmonic fields from Eq.~(1).

In Fig.~1(c), the computed HHG spectra from our model are compared with PIC simulations for various laser normalized amplitudes.
Notably two distinct groups of harmonic beams are generated, the odd- and even-order harmonics are polarized orthogonal to each other. 
In both cases, the harmonics spectra have a power-law shape $I_n \propto n^{\alpha}$, with the index $\alpha$ increases with $a_0$ until it saturates around $\alpha\approx-3.5$. 
This can be interpreted as the increasing of peripheral electron oscillation amplitude is limited by causality,
namely, $\delta_2$ must be smaller than the maximum distance an electron can travel within half of the oscillating period. 
Substituting $\delta_2=0.25\lambda_0$ into the model, the resulted harmonic spectrum is presented by the black circles in the Fig~1(c), which agree very well with the PIC simulation.

However, since the peripheral electron oscillate with $2\omega_0$, it can only explain the odd-order (normal) harmonic beams.
The generation of even-order (anomalous) harmonics requires a zero-frequency field to be present at the oscillating slit.
This condition is fulfilled by the charge separation between electrons and heavy ions, which arises from the laser ponderomotive force pushing electrons into plasma. 
As a result, this electrostatic field is directed perpendicular to the slit, and attached to a boundary that oscillating at $2\omega_0$.
Therefore, even-order harmonics emerge, whose polarization are naturally orthogonal to the driving pulse.
We show that by adding an electrostatic field in Eq.~(1), the anomalous harmonics can be reproduced from our model, as shown by the black squares in Fig.~1(c). Notably, they follow the same power-law spectral slope as the odd-order harmonics, but with slightly weaker overall strength.\\

\begin{figure}[!t]
	\centering
	\includegraphics[width=8.5cm]{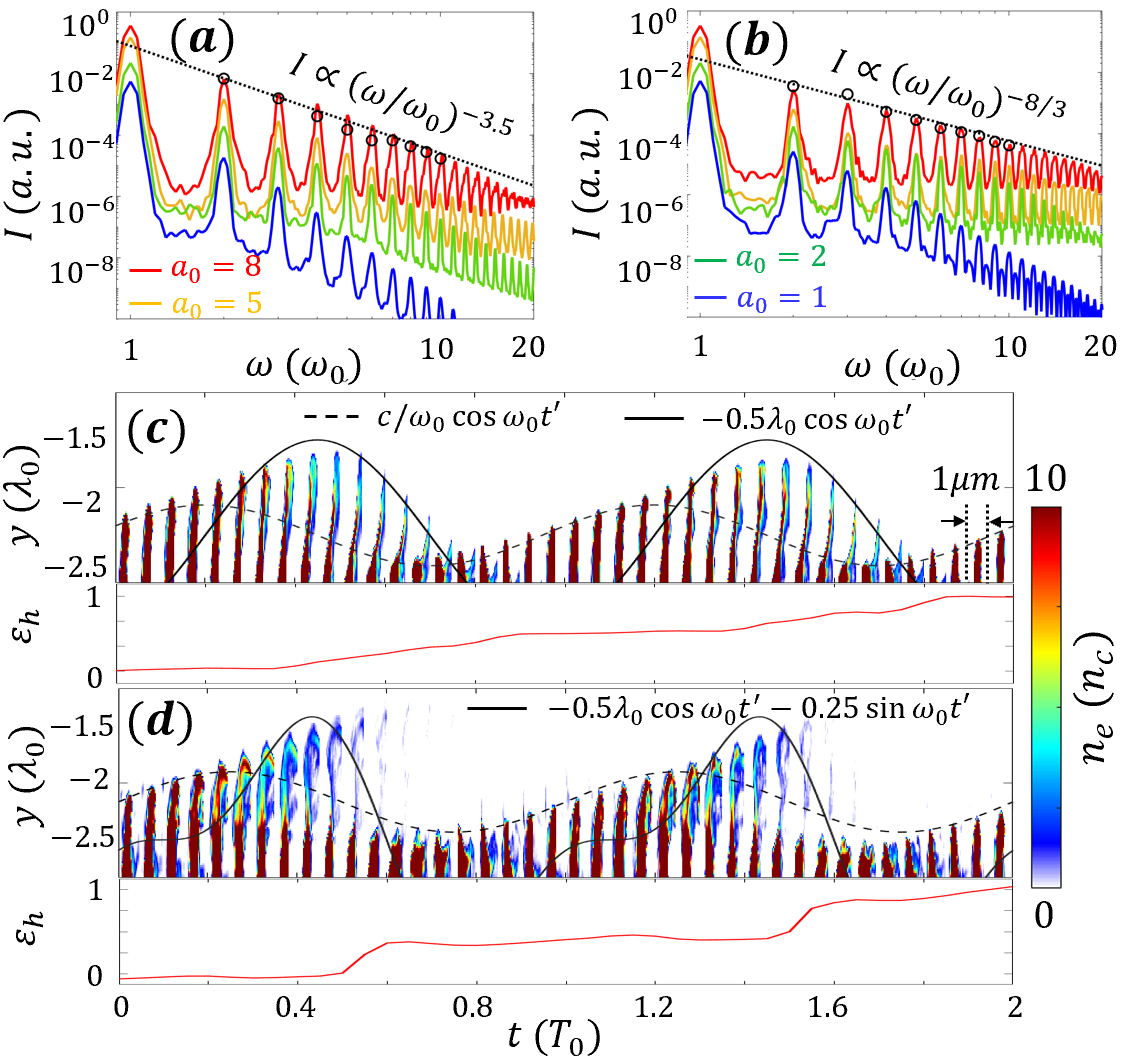}
	\caption{The HHG spectra from the (a) CP and (b) LP$_\perp$ drivers, where the solid lines show the results from PIC simulations with color representing the normalized laser amplitude, and the black circular makers show the saturated spectrum calculated from Eq.~(1). The upper panels of (c) and (d) shows the low-energy $(\gamma<3)$ electron density oscillation at the lower boundary of slit (corresponding to the blue rectangular in Fig.~1(a)) for the CP and LP$_\perp$ drivers, respectively, where each column represents a snapshot of $n_e$ within $x = x_0\pm 0.5 \mu m$ temporally spaced by $0.05 T_0$. The black solid and dashed curves show the fitting (specified in the corresponding plot) we used for calculating the harmonic fields. The lower panels of (c) and (d) show the harmonic energy (with $\omega\geq2\omega_0$) evolution over the same period of time.}
\label{fig:2}
\end{figure}

With the nonlinear effect of ponderomotive force been correctly addressed by the model, we can then discuss the 
saturated power-law HHG spectral shape $I_n\propto n^{-3.5}$ (the ``SWB limted spectrum" hereafter).
As shown for the LP$_\parallel$ driver above, and also seen in the CP case (Fig.~2(a)), such spectrum commonly arises when: (i) the peripheral electron oscillation can be estimated by a {\it harmonic oscillation} during SWB phase; and (ii) its amplitude is limited by causality.
However,  as shown in Fig.~2(b), the saturated HHG spectrum in the LP$_\perp$ case is slightly slower, with a power-law index $\alpha\approx-8/3$.
This can be attributed to the violation of (i) due to ponderomotive effects.

The peripheral electron dynamics are compared in detail for the CP can LP$_\perp$ drivers in Figs.~2(c-d), where upper panels depict electron density evolution at lower boundary of the slit ($y<0$), and the lower panels show the harmonic energy increasing over the same period of time.
Apparently, when the electrons oscillate towards the plasma bulk, the temporal trajectory of slit boundary roughly follows $d\mathbf{R'}(t') = -c/\omega_0\cos(\omega_0 t')\mathbf{e}_y$ (dashed lines) for both cases, and the HHG during this phase is negligible. 
On the other hand, as the electrons oscillate towards vacuum, periodic outbursts of SWB are observed. This is when the harmonic fields are predominantly generated. 
Importantly, the peripheral electron dynamics in two cases deviate during this phase.
For the CP drivers, there is no oscillating terms in the ponderomotive force, the SWB electron dynamics can be estimated by a harmonic motion $d\mathbf{R'}(t') = -0.5\lambda_0\cos(\omega_0 t')\mathbf{e}_y$ (black solid line in Fig.~2(c)). 
Our model thus yield the SWB limited spectrum, which agrees very well with PIC simulations (Fig.~2(a)).
While for the LP$_\perp$ drivers, the electron dynamics 
cannot be modeled by a harmonic motion due to the $2\omega_0$ oscillating term in the ponderomotive force, and the resulted HHG spectrum surpass the SWB limit.

Nevertheless, our model can be easily modified to account for such nonlinear oscillations. By adding a second-harmonic term, namely $d\mathbf{R'}(t') = [-0.5\lambda_0\cos(\omega_0 t') \pm 0.25\lambda_0\sin(2\omega_0 t')]\mathbf{e}_y$ (black solid curve in Fig.~2(d)), where the $\pm$ sign corresponding to the $\pm y$ side of the slit, our model is able to  reproduce the saturated HHG spectrum for the LP$_\perp$ case, as shown by the black circles in Fig.~2(b).

Notably, the HHG spectrum produced by a relativistic LP$_\perp$ drive pulse diffracting through a single slit becomes almost the same as the ROM scenario \cite{Baeva2006}. This is because when a high power laser obliquely incident on a reflective plasma foil, the oscillating electron layer follows a similar nonlinear motion as discussed in \cite{Lichters1996}.
However, there is a small difference, unlike the ROM being a quasi-1D model \cite{Lichters1996}, the diffracted lights are spatially inhomogeneous.
The HHG spectra reported in this work are obtained by integrating over all diffracting angles, this might be responsible for the discrepancy with the results reported in Ref.~\cite{King2023}.\\

Finally, we discuss another anomalous effect in the relativistic single-slit diffraction. It sterns from the total angular momentum conservation in the HHG process driven by a CP pulse. 
When $N$ laser photons are converted into one $N$th harmonic photon during the diffraction, their (spin) angular momenta $N\sigma_z\hbar$ must be transferred to the harmonic photon generated, where $\sigma_z = 1$ and $-1$ corresponding to right-handed and left-handed circular polarization, respectively. 
Previous study \cite{Yi2021} shows this effect converts the spin angular momentum of the drive CP pulse into the intrinsic OAM of the harmonics in relativistic pinhole diffraction. However, this process requires rotational symmetry, thus cannot occur in single-slit diffraction. 

\begin{figure}[!t]
	\centering
	\includegraphics[width=8.5cm]{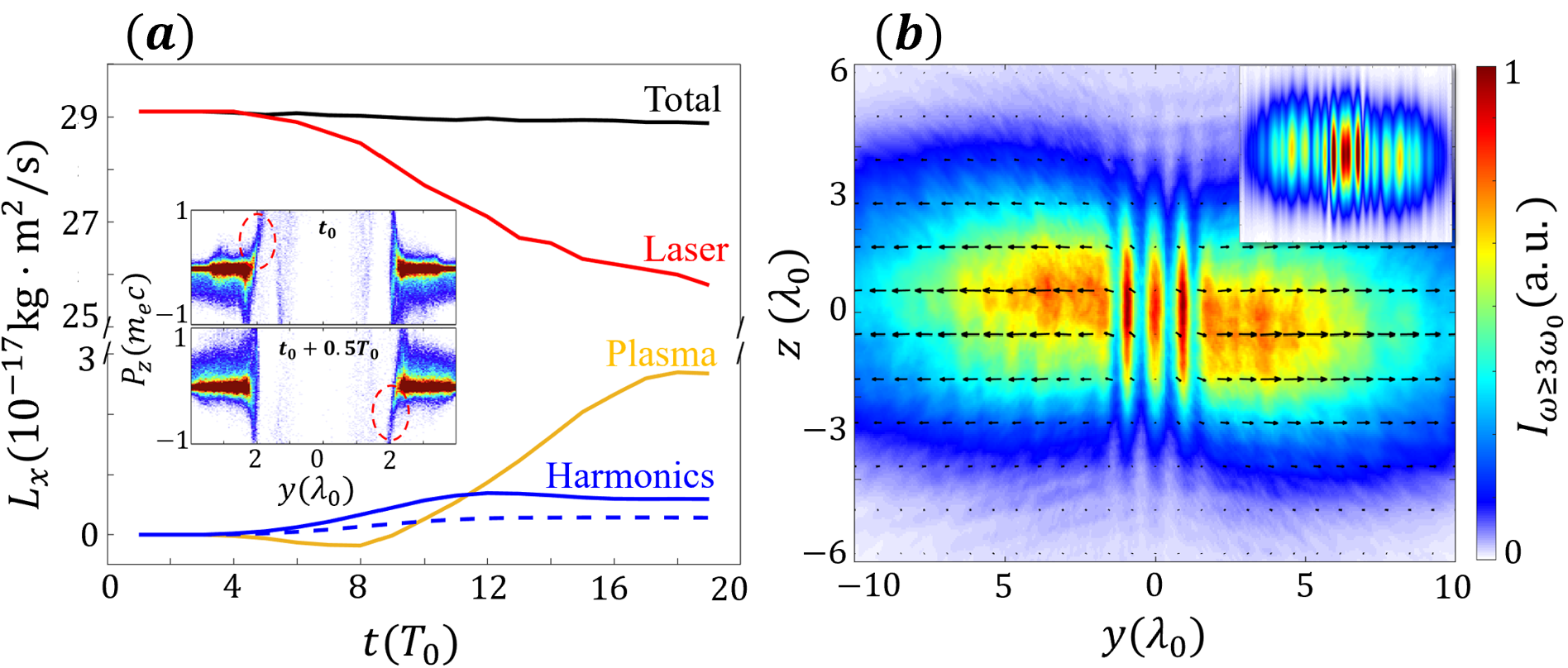}
	\caption{(a) The temporal evolution of total angular momentum aligned with beam axis carried by each components in the 3D PIC simulations are presented by the solid curves, and the blue dashed line shows the spin angular momentum of the harmonic beams estimated by their total energy. The inset of (a) shows phase space diagram $(y-P_z)$ for peripheral electrons at two times separated by $0.5T_0$, where the red dashed circles highlight the electrons undergoing SWB process. (b) The tilted near-field intensity pattern of harmonic fields ($\omega\geq3\omega_0$), where the black arrows represent the transverse momentum of the harmonics, and the inset shows the prediction from our model.}
	\label{fig:3}
\end{figure}

The answer to this dilemma lies in the extrinsic OAM gained by the harmonics due to the tangential electron motion with respect to the boundary of the slit. When the peripheral electrons move towards vacuum during the SWB phase, the conservation of transverse canonical momentum leads to $\mathbf{P_\perp} = -(\hat{e}_y+i\sigma\hat{e}_z)a_0\exp{(-i\phi_{swb})}$, where 
$\phi_{swb}$ is the optical phase that SWB effect occurs. 
This indicates a tangential shift of the peripheral electrons as they oscillates, which is controlled by the polarization of drive pulse.
For a right-hand drive CP pulse, the peripheral electrons simultaneously streaming towards $+z$ ($-z$) when they oscillate inwards at the $-y$ ($+y$) side of the slit, this is verified by the phase space map ($y-P_z$) shown in the inset of Fig.~3(a).

Fig.~3(a) illustrates the temporal evolution of the angular momenta that align with the beam axis carried by each component during the diffraction process. One can see the fundamental laser losses around $10\%$ of its (spin) angular momentum, which are transferred to the plasma and harmonic beams, the total angular momentum of the system conserves.
Notably, as indicated by the phase space diagram shown in the inset, it is expected that the plasma picks up a negative angular momentum anti-parallel to the drive pulse because of the tangential shift, this is observed during the diffraction process for $6T_0<t<10T_0$. 
In addition, we estimate the spin angular momentum by $L_s \approx E_n/(n\omega_0)$ for each harmonic order $n$, and found their sum (blue dashed line) is much smaller than their total angular momenta.
This indicates the harmonic beams are carrying significant (extrinsic) OAM.

Fig.~3(b) depicts the near-field harmonic intensity pattern ($n>3$), with the transverse electromagnetic momenta (black arrows) superposed to the plot.
Obviously, an extrinsic OAM align with $+x$ arises due to the tilted harmonic intensity pattern. This ensures total angular momentum conservation for the HHG process, akin to the Imbert-Fedorov shift in non-relativistic optics \cite{Bliokh2013}.
In order to explain this effect, one needs to consider a drive pulse with finite focal spot. Thus the tangential shift of peripheral electrons leads to the deformation of the (effective) diffraction window. 
Fortunately, this effect can now be investigated with the present model. 
By submitting
$d\mathbf{R}(t') = -0.5\lambda_0(\hat{e}_y+i\sigma\hat{e}_z)\exp(\omega_0 t')$ into Eq.~(1), the calculated harmonic intensity pattern (inset of Fig.~3(b)) is in reasonable agreement with the 3D PIC simulations.
We noted that the tangential motion of electrons does not change the harmonic spectrum, it only modifies the intensity pattern due to conservation of total angular momenta.\\

In conclusion, we have developed a theoretical model for relativistic diffraction, that allows for calculating HHG fields from arbitrary 2D electron motion within the diffraction plane.
Using this model, we study a few anomalous effects in the relativistic single-slit diffraction: (i) the generation of anomalously polarized harmonic beams for LP$_\parallel$ drivers, which can be attributed to a combined effect of ponderomotive force and charge separating field at the plasma-vacuum interface; (ii) explain the harmonic spectrum slope beyond the SWB limit for the LP$_\perp$ case; (iii) a Imbert-Fedorov-like shift observed for CP laser-driven harmonic beams, where extrinsic OAM arises due to tangential peripheral electron motion.
The presented model provides a general solution to any non-planar wave diffraction problem, which is critical for interpreting complex electromagnetic wave diffraction, such as Laguerre Gaussian beams \cite{Allen1992,Shi2014}, vector light fields \cite{RG2018,Chen2022}, and spatial temporal optical vortices \cite{Bliokh2015pr}. These are discussed in other works\cite{Hu2025}.\\


\begin{acknowledgments}
	This work is supported by the National Key R$\&$D Program of China (No. 2021YFA1601700), and the National Natural Science Foundation of China (No. 12475246).
\end{acknowledgments}

\end{document}